\documentclass[paper]{agujournal2019}
\usepackage{url} 
\usepackage{lineno}
\usepackage{soul}
\usepackage{xcolor}
\usepackage{tabularx}
\usepackage{amsmath}
\usepackage{siunitx}
\usepackage{cleveref}
\usepackage{makecell}

\usepackage{multirow}
\newcommand{\tenso}[1]{\mathrm{\boldsymbol{#1}}}

\DeclareMathOperator\erf{erf}
\UseRawInputEncoding

\definecolor{gauthiercolor}{HTML}{037D50}

\journalname{Water Resources Research}

\begin{document}
\def\thefootnote{*}\footnotetext{These authors contributed equally to this work}\def\thefootnote{\arabic{footnote}}


\title{Dispersion versus diffusion in mixing fronts}

\authors{Gauthier Rousseau$^*$\affil{1,2}, Satoshi Izumoto$^*$\affil{1}, Tanguy Le Borgne\affil{1} and Joris Heyman\affil{1}}

\affiliation{1}{Univ. Rennes 1, CNRS, G\'eosciences Rennes, UMR 6118, 35000 Rennes, France}
\affiliation{2}{Institute of Hydraulic Engineering and Water Resources Management, TU Wien, Karlsplatz 13, 1040, Vienna, Austria}

\correspondingauthor{Joris Heyman}{joris.heyman@univ-rennes.fr}

\begin{keypoints}
\item The effect of diffusive versus local dispersion on mixing front dynamics is studied under uniform and non-uniform flows
\item Analytical solutions for the mixing width and flux are derived for all considered scenarios and validated through simulations and experiments 
\item While diffusion and local dispersion produce similar mixing front dynamics in uniform flows, they fundamentally differ in non-uniform flows
\end{keypoints}

\begin{abstract}
Mixing fronts form when fluids with different chemical compositions are brought into contact. They influence a large range of biogeochemical processes in hydrological systems. 
An important mechanism governing mixing rates in such fronts is stretching by non-uniform flows that accelerates diffusive mass transfer by enhancing concentration gradients. In a range of systems, including porous media at Darcy scale, hydrodynamic dispersion dominates over diffusion to control local mixing rates. As it differs from diffusion through its velocity-dependent dispersion tensor, it is not known how local dispersion interacts with  macroscopic mixing front stretching. 
Here, we investigate the impact of local dispersion versus diffusion on the properties of steady mixing fronts created by both uniform and non-uniform flows. We derive analytical solutions for the concentration profile, mixing scale and mixing rate across the fronts. 
We validate these predictions by comparison with numerical simulations and experiments performed in quasi two-dimensional tanks over a broad range of P\'eclet numbers. 
Without porous media, the mixing scale is governed by local diffusion coupled with flow: it increases diffusively along streamlines in uniform flows while it is constant in converging flows due to the balance between fluid compression and local diffusion. 
With porous media, the Batchelor scale is no longer sustained and the mixing scale grows with dispersion in non-uniform flows. In addition,  the coupling between flow acceleration and dispersion results in a P\'eclet independent mixing interface, in contrast with the local diffusion scenario. We discuss the consequences of these findings on mixing rates in mixing fronts.

\end{abstract}

\section*{Plain Language Summary}

This study explores how mixing in fluids with and without the presence of a dispersive matrix such as porous media.  It highlights the difference of mixing rates in miscible solutes interfaces in the presence or absence of hydrodynamic dispersion, in uniform and non-uniform flow fields. We find that the mixing scales and mixing rates in the presence of dispersion media follows different laws than their well-known diffusive counterparts. These findings provide new insights into mixing and reaction processes controlling a range of applications, such as contaminant transport and remediation, water quality, and subsurface energy storage and extraction.

\section{Introduction}
Mixing fronts formed by miscible fluids influence a range of hydrological and biogeochemical processes \cite{Dentz2011,Rolle2019,Valocchi2019b} including river-groundwater exchanges\cite{Hester2017,Bandopadhyay2018ShearHillslopes,Ziliotto2021MixingExperiments}, freshwater-saltwater mixing in coastal areas~\cite{abarca2007anisotropic,deVriendt2021},  subsurface microbial processes \cite{Bochet2020} and mixing in river confluences and estuaries \cite{prandle2009estuaries,bouchez2010turbulent, yuan2022dynamics}. They are also present in many engineering applications, such as 
soil and groundwater remediation ~\cite{Karadimitriou2012a,Fu2014,Wang2022EnhancedExtraction},  geological carbon sequestration~\cite{Zoback2012, Szulczewski2012LifetimeTechnology}, hydrogen storage \cite{Tarkowski2019UndergroundProspects,Lysyy2022HydrogenStorage} and geothermal systems~\cite{Burte2019KineticReactions}.
The heterogeneity of flow fields at various scales leads to stretching of mixing fronts, leading to the enhancement of concentration gradients and the resulting mixing rates ~\cite{rolle2009enhancement, cirpka2011stochastic,de2012flow, chiogna2012mixing,  LeBorgne2014, engdahl2014predicting, cirpka2015transverse, ye2015experimental, Rolle2019,Villermaux2019,Ye2020,dentz2023mixing}. The interplay between fluid stretching and diffusion has been captured by lamellar models, that represent mixing fronts as stretched elementary lamellae (in 2D) or sheets (in 3D), in porous media ~\cite{LeBorgne2013,LeBorgne2015,Heyman2020, Souzy2020} and turbulent flows~\cite{Villermaux2019}. 
While this approach has successfully described the coupling between stretching and diffusion, it does not account for situations where dispersion dominates locally over diffusion. This occurs in a range of hydrological systems, including porous media at Darcy scale  ~\cite{Dentz2011}.

Hydrodynamic dispersion is driven by the combination of molecular diffusion and pore-scale flow variability~\cite{Bear1988}. It is classically modeled as an anisotropic effective dispersion coefficient, whose magnitude is proportional to flow velocity~\cite{Delgado2007}, as opposed to constant diffusion coefficients. 
Such velocity dependent local dispersion is disregarded in studies that use Hele-Shaw cells as experimental analogues of porous media to study mixing fronts, density driven flows, and reactions~\cite{de2020chemo}
 The velocity dependence of the dispersion coefficient has been shown to play an important effect on mixing rates at the salt-fresh water interfaces in porous media \cite{de2022subsurface} and in Poiseuille flows~\cite{perez2019upscaling}.
 However, it is not generally known how to quantify its effect on mixing front properties under non-uniform flows.  
 
Here, we investigate the role of dispersion versus diffusion on mixing front properties under both uniform and non-uniform flows. As a paradigm of non-uniform flow, we consider a mixing front created by the convergence of opposing flows. Such flow leads to the appearance of a stagnation point (a point of null velocity) and frequently arise in hydrological systems \cite{Bresciani2019}. This includes hyporheic zones where groundwater upwelling locally competes with flow recirculation produced by variations in the river bathymetry \cite{Hester2017}, fresh-salt water interfaces \cite{deVriendt2021},  density-driven flows \cite{Hidalgo2015}. The velocity field close to a stagnation point is non-uniform, with a constant velocity gradient magnitude close to the stagnation point. This means that the flow is constantly decelerating/accelerating when approaching/departing from the stagnation point. This leads to a net stretching of fluid elements and a constant compression rate of mixing fronts. Under the assumption of local diffusion, such compression is known to sustain chemical gradients over a fixed characterize length scale called the Batchelor scale~\cite{Villermaux2019}.
While the coupling between stretching and diffusion is well understood is such flows, it is not known how these dynamics are altered by local dispersion. 

The paper is organized as follows. In the first section, we solve analytically the steady advection-dispersion equation governing a conservative mixing interface under a constant compression rate. We obtain analytical predictions for the concentration profile, the mixing scale and the mixing rate across the interface.  We validate these predictions by comparison with numerical simulation of the coupled flow-dispersion problem. In the second section, we compare these predictions to tank experiments of conservative tracer in uniform and non-uniform flows. We investigate mixing fronts both in the presence and absence of porous media, which lead to respectively local diffusion and local dispersion.

\section{Theory}
\label{sec:Theo}

\def\H{2.3 cm}

\subsection{Mixing under advection diffusion/dispersion}
\label{sec:flowConf}

\begin{figure}
    \includegraphics[width=\linewidth]{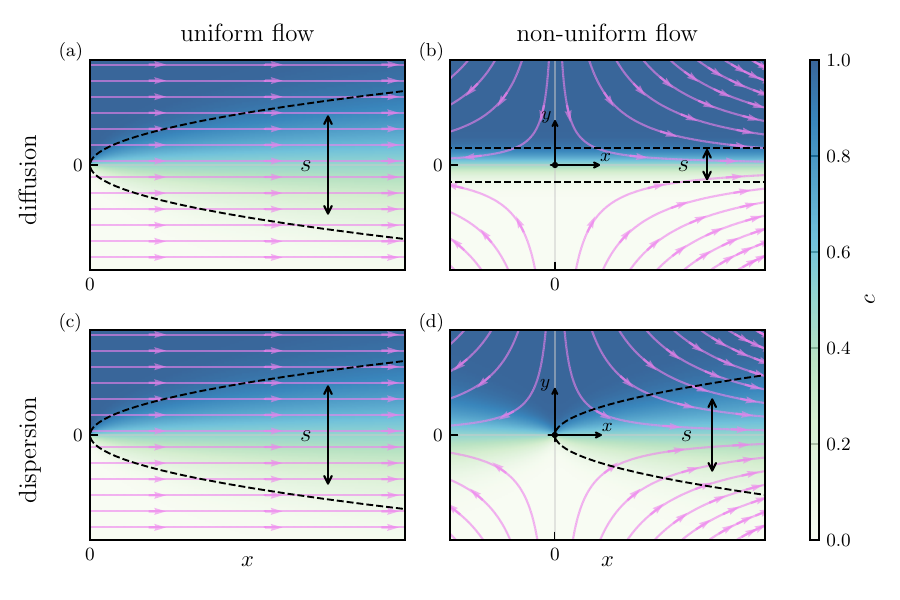}
    \caption{ 
    Numerical simulations showing the mixing interface of a conservative solute $c$ in a uniform (left column) and non-uniform flow (right column), if dispersion is assumed constant (top) or proportional to the velocity (bottom). The dotted lines show the extent of mixing width given by theory (\cref{eq:coWidth} for a and c, $s_B= \sqrt{ 2 {D} / \gamma} $ for b and \cref{eq:dispSaddleWidth} for d.)} 
    \label{fig:sketch}
\end{figure}

We consider the transport of a conservative solute in a two-dimensional incompressible flow field with velocity $\boldsymbol{u}(x,y)$, following:

\begin{equation}
  \frac{\partial c}{\partial t} = -\boldsymbol{u}(x,y) \cdot  \boldsymbol{\nabla}  c(x,y) + \nabla \cdot (\tenso{D}(x,y)\, \cdot \boldsymbol{\nabla} c(x,y) ),
    \label{eq:advdisp}
\end{equation}
where $\tenso{D}$ is the dispersion tensor. 
To compare the effect of diffusion and dispersion, we consider two models: (i) a constant diffusion coefficient, $\tenso{D}(x,y)=D \tenso{I}$ or (ii) a velocity dependent dispersion coefficient \cite{Bear1988}:
\begin{equation}
\tenso{D} = \left(D_m + \alpha_T |\boldsymbol{u}| \right) \tenso{I}+ \left(\alpha_L-\alpha_T\right) \frac{\left(\boldsymbol{u}  \otimes \boldsymbol{u}\right)}{|\boldsymbol{u}|}
\label{eq:disp}
\end{equation}
where  $|\boldsymbol{u}|$ denotes the norm of the velocity vector, $\otimes$ is the outer product, $\tenso{I}$ is the identity matrix and $D_\text{m}$ the effective molecular diffusion in porous media. $\alpha_T$  and $\alpha_L$ are the transverse and longitudinal dispersivities. 
\Cref{eq:disp} implies that the dispersion coefficients parallel and transverse to the local flow direction are respectively
\begin{equation}
{D}_\text{L}={D}_\text{m}+\alpha_L |\boldsymbol{u}|
\label{eq:disp_long}.
\end{equation}
and
\begin{equation}
{D}_\text{T}={D}_\text{m}+\alpha_T |\boldsymbol{u}|.
\label{eq:disp_trans}
\end{equation}
To highlight the difference produced between these two models in mixing front properties, we consider a uniform and a non-uniform flow field. We define the P\'eclet number as 
\begin{equation}
\text{Pe}=
\frac{U d}{D_\text{m}},
 \label{eq:peclet}
\end{equation}
where $U$ and $d$ are respectively a characteristic velocity and  grain size.

\subsection{Uniform flow}
First, we consider the trivial case of a uniform flow with constant velocity $U$ (\cref{fig:sketch}.a) where
 \begin{equation}
            \boldsymbol{u}(x,y)= \left[ \begin{array}{c}
                U \\
                0 
            \end{array}\right].
            \label{eq:velcoflow}
\end{equation}
Since the flow velocity is constant, dispersion is also a constant tensor $\tenso{D}$ and the transport equation is
\begin{equation}
 \frac{\partial c}{\partial t} = -\boldsymbol{u}\cdot \boldsymbol{\nabla} c + \tenso{D} \Delta  c
 \label{eq:CD}
\end{equation}
We impose the boundary condition to be a continuous solute injection at the flow inlet on the half plane ($c(x,y>0)=1$), a so-called co-flow configuration (\cref{fig:sketch}.a). Continuous time-independent injection leads to the existence of a steady-state solution of the mixing interface, so that $\partial c/\partial t=0$. As illustrated in \cref{fig:sketch}.a and \cref{fig:sketch}.c, concentration gradients are much larger along $y$ than along $x$ directions, and mixing occurs mostly through transverse diffusion/dispersion.
Hence, \cref{eq:CD} can be approximated by:
\begin{equation}
U\frac{\partial c}{\partial x} =  D_\text{T} \frac{\partial^2 c}{\partial y^2} 
\label{eq:diffco}
\end{equation}
whose solution is
\begin{equation}
c(x,y) = 1/2 \left[ 1  + \erf \left( \frac{y}{ s(x)} \right) \right]
\label{eq:err}
\end{equation}
where $\erf$ is the error function. The mixing width $s(x)$ is:
\begin{equation}
s = 2\sqrt{\frac{D_\text{T} }{U}x}.
\label{eq:coWidth}
\end{equation}
The characteristic concentration gradient and mixing flux along the interface are respectively,
\begin{equation}
\nabla c (x) \sim \frac{1}{s} \sim \sqrt{\frac{U}{{D_\text{T} } x}}
\label{eq:grad}
\end{equation}
and
\begin{equation}
J(x)= D_\text{T}  \nabla c \sim \sqrt{\frac{U D_\text{T} }{x}}
\label{eq:flux}
\end{equation}
The mixing front properties for diffusion and dispersion dominated regimes are thus obtained by substituting  $D_\text{T}$ from equation \eqref{eq:disp_trans} in the equations above. The results are synthesized in table \ref{tab:width}.
\subsection{Non-uniform flow}
We now focus on the case of a non-uniform flow characterized by a constant stretching/compression rate $\gamma$ induced by converging flows (\cref{fig:sketch}.b). This flow represents a paradigm of stretching enhanced mixing \cite{Villermaux2019,Rolle2019, Izumoto2022}.
Linearization of the flow field around the stagnation point leads to the velocity field
        \begin{equation}
            \boldsymbol{u}(x,y)= \left[ \begin{array}{cc}
            \gamma & 0 \\
            0 & - \gamma \end{array}\right] \left[ \begin{array}{c}
             x \\
             y \end{array}\right] + o(x^2,y^2,xy),
            \label{eq:velsaddle}
        \end{equation}
where $\gamma=\partial \boldsymbol{u} / \partial x |_{0,0}$. This velocity field imparts a constant compression rate $\gamma^{-1}$ to fluid elements in the $y$ direction, and a constant stretching rate $\gamma$ in the $x$-direction. Compression acts at enhancing scalar gradients in the $y$ direction, with a direct impact on solute transport and mixing as discussed below. 
To define the P\'eclet number (equation \eqref{eq:Peclet}), we take the characteristic velocity $U=\gamma L$, where $L$ is the macroscopic length scale over which the stagnation point develops.

\subsubsection{Diffusion in non-uniform flow}
We first consider a constant diffusion coefficient isotropic in the flow domain $\tenso{D}=D_\text{m} \tenso{I}$. Since the compression rate is constant, this system is locally equivalent to chaotic flows where constant compression/stretching rates are sustained globally due to exponential elongation rates~\cite{Batchelor1959,Lester2013, Lester2016, Villermaux2019, Heyman2020}. The balance of compression rate of fluid elements $\frac{1}{s}\frac{d s}{d t}=-\gamma$  and the rate of diffusive expansion $\frac{1}{s}\frac{d s}{d t}=\frac{2 D_\text{m}}{s^2}$ leads to the emergence of a fixed mixing scale
\begin{equation}
s_B\sim \sqrt{ \frac{2 D_\text{m}}{\gamma}}.
\label{eq:batchelor}
\end{equation}
This scale, at which solute gradients are maintained, is called the Batchelor scale \cite{Batchelor1959, Villermaux2019}. 
The concentration profile across the mixing interface is obtained by solving \cref{eq:advdisp} with the linearized flow field (\cref{eq:velsaddle}). The boundary conditions consist of continuous injection of solute on one side of the stagnation point, e.g. $c(x,y=-\infty)=1$ and $c(x,y=\infty)=0$. The solute $c$ is advected and mixes transversally along the mixing interface located on the $x$-axis (\cref{fig:sketch}b). Concentration gradients along $x$ are null because both the $y$ component of the velocity and the boundary conditions are independent of $x$. As before, the continuous injection of solute leads to a steady-state mixing interface, govern by:
\begin{equation}
-  y \frac{\partial c}{\partial y}   = \frac{{D_\text{m}}}{\gamma} \frac{\partial^2 c}{\partial y^2}.
\label{eq:diffSF}
\end{equation}
The solution of this equation is an error function,
\begin{equation}
c(x,y) = 1/2 \left[ 1  + \erf \left( \frac{y}{ s_B} \right) \right]
\label{eq:err_sB}
\end{equation}
with a constant mixing width given by the Batchelor scale $s=s_B$.
The characteristic concentration gradient and mixing flux per unit area of the interface are respectively,
\begin{equation}
\nabla c (x) \sim \frac{1}{s_B} \sim \sqrt{\frac{\gamma }{ D_\text{m}}} 
\label{eq:grad}
\end{equation}
and
\begin{equation}
J(x)= D_\text{m}\nabla c \sim \sqrt{D_\text{m}\gamma}  
\label{eq:flux}
\end{equation}
We synthesize the results in table \ref{tab:width}.

\subsubsection{Dispersion in non-uniform flow}
\label{seq:dispmix}
We now consider a dispersion coefficient proportional to flow velocity, and thus varying in space. We impose the same boundary conditions as before and assume that the mixing interface remains close to the position $y=0$, where the flow velocity is almost parallel to the $x$ direction. Within such approximation, \cref{eq:advdisp} simplifies to
\begin{equation}
\frac{\partial c}{\partial t} =  \gamma \,y \frac{\partial c}{\partial y} - \gamma \,x \frac{\partial c}{\partial x} - \frac{\partial}{\partial y} \left( (D_m+\alpha_T \gamma \, x)  \frac{\partial c}{\partial y}\right) + \frac{\partial}{\partial x} \left( (D_m+\alpha_L \gamma \, y) \frac{\partial c}{\partial x}\right) .
\label{eq:fullsaddle0}
\end{equation}
We focus on the solution away from the stagnation point, where $y/x = \delta \ll 1$ and $\alpha_T \gamma x \gg D_m$.  We thus ignore the contribution of molecular diffusion in \cref{eq:fullsaddle0}.

We define new dimensionless variables $(X,Y)=(x/L_x,y/L_y)$, with $L_x$ and $L_y$ being the characteristic length scales of variation of concentration in $x$ and $y$ directions. \Cref{eq:fullsaddle0} becomes:
\begin{equation}
    - \delta_{L}^2 \,Y \frac{\partial c}{\partial Y} + \delta_{L}^2\, X \frac{\partial c}{\partial X} = \frac{\alpha T}{L_y}\frac{\partial}{\partial Y} \left(  X  \frac{\partial c}{\partial Y}\right) + \frac{\alpha L}{L_x} \delta_{L}^2 \frac{\partial}{\partial X} \left(   Y \frac{\partial c}{\partial X} \right)
    \label{eq:fullsaddle3}
\end{equation}
where $\delta_L=L_y/L_x$. Noting that the variations of concentration transverse to the mixing interface are more pronounced than longitudinal ones, we assume $\delta_L\ll 1$ and $\alpha_T/L_y \gg \alpha_L/L_x$. Keeping only the leading order terms in $\delta$, \cref{eq:fullsaddle0} simplifies to
\begin{equation}
- \gamma \,y \frac{\partial c}{\partial y} + \gamma \,x \frac{\partial c}{\partial x} = \frac{\partial}{\partial y} \left( \alpha_T \gamma \, x  \frac{\partial c}{\partial y}\right).
\label{eq:fullsaddle1}
\end{equation}
The concentration profile is found by assuming the functional form $c(y)= A \, \erf \left(\frac{y}{\sqrt{K \alpha_T x}} \right) + B$. $K=4/3$ is determined so that the equation satisfies \cref{eq:fullsaddle1}, and boundary conditions. Thus, the solution is,
\begin{equation}
   c(y) = 1/2 \left[ 1  + \erf \left( \frac{y}{\sqrt{4/3 \alpha_T x}} \right) \right]
   \label{eq:solution_dispersion}
\end{equation}
This leads to a mixing width, 
\begin{equation}
   s=2\sqrt{\frac{\alpha_T x}{3}}.
   \label{eq:dispSaddleWidth}
\end{equation}
The characteristic concentration gradient and mixing flux per unit area of the interface are respectively,
\begin{equation}
\nabla c (x) \sim \frac{1}{s} \sim \sqrt{\frac{3 }{4\alpha_T x}} 
\label{eq:grad}
\end{equation}
and
\begin{equation}
J(x)= \alpha_T \gamma x \nabla c \sim  \gamma  \sqrt{\frac{3 \alpha_T x}{4}}
\label{eq:flux}
\end{equation}
We validated the solution \eqref{eq:dispSaddleWidth} by numerical simulations of the full advection-dispersion equation using the OpenFOAM code (\cref{fig:sketch} and Appendix~\ref{sec:valid}).
Thus, the mixing front properties are similar as for uniform flows, although the mixing width is reduced by a factor $1/\sqrt{3}$.   We summarizes the theoretical dependence of the mixing width and mixing rate with space and P\'eclet number in Table~\ref{tab:width}. 

   \begin{table}
\begin{center}
        \begin{tabular}{ |c|c|c|  }
            \hline
           & uniform flow & non-uniform flow \\ \hline
           \multicolumn{3}{|c|}{Mixing width $s$} \\ \hline
           diffusive  &  $s \sim x^{1/2}  {\text{Pe}}^{-1/2}$ & $ s \sim x^0  {\text{Pe}}^{-1/2}$  \\ \hline
           dispersive & $ s \sim x^{1/2} {\text{Pe}}^{0}$ & $ s \sim x^{1/2} {\text{Pe}}^{0}$\\ \hline
       
           \multicolumn{3}{|c|}{Mixing rate $J$} \\ \hline
        
           diffusive  &  $ J \sim x^{-1/2}  {\text{Pe}}^{1/2}$ & $ J \sim x^0  {\text{Pe}}^{1/2}$  \\ \hline
           dispersive & $ J \sim x^{1/2} {\text{Pe}}^{1/2}$ & $ J \sim x^{1/2} {\text{Pe}}^{1}$\\ \hline
        \end{tabular}
\caption{Theoretical mixing width $s$ and rate $J$ dependence with spatial coordinate along the mixing interface $x$ and P\'eclet number (equation \eqref{eq:peclet}) for the two flow conditions.}\label{tab:width}
\end{center}
 \end{table} 

\section{Experiments}
\begin{figure}
    \centering
    \includegraphics[width=\linewidth]{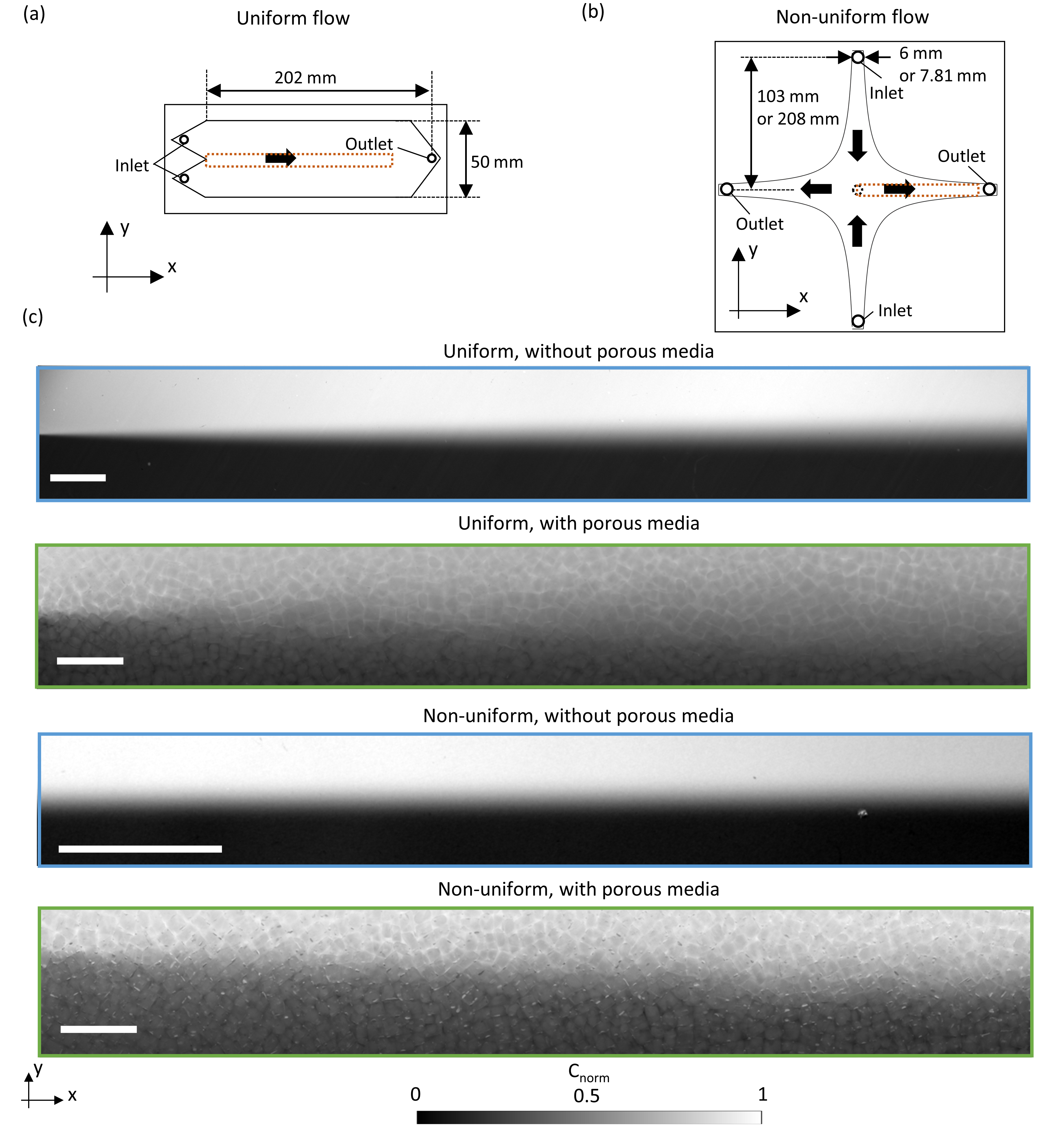}
    \caption{(a,b) Experimental design of (a) the uniform flow cell and (b) non-uniform flow cell. The brown dotted lines indicate the flow regions shown in (c). The thickness of cell is 2mm and 12mm without and with porous media respectively. The non-uniform cell has a length of 103mm ($a=303$ mm$^2$) without porous media and 208mm with porous media ($a=811$ mm$^2$). The thick arrows indicate the direction of the flow. (c) Tracer concentration field in uniform flow (top two rows) and non-uniform flow cases (bottom two rows). Flow is from left to right. The white bars represent 10 mm. }
    \label{figexpImages}
\end{figure}

In the following, we test the theoretical predictions against conservative tracer experiments in quasi-two dimensional cells in the presence and absence of porous media. 

\label{sec:protocol}
\subsection{Experimental setup}
Two Hele-Shaw cells were designed to achieve (i) uniform flow and (ii) non-uniform flow created by converging flows. The thickness of the cells was chosen to be small compared to the longitudinal and transverse dimensions. The uniform flow cell is a rectangular cell with a co-injection of tracers at the inlet (\cref{figexpImages}a). The co-injection of two solutions produces a mixing interface that propagate towards the outlet at the other side of the cell (\cref{figexpImages}a). The non-uniform flow cell is designed with four branches of hyperbolic shape with $y = \pm a/x$, $a$ being a constant, that reproduces the geometry of streamlines near a stagnation point created by opposing flows (equation \eqref{eq:velsaddle}). Flow inlets and outlets are facing each other (\cref{figexpImages}b), creating a stagnation point and an horizontal mixing interface in the middle cell.  

Two sets of conservative transport experiments were performed in these Hele-Shaw cells, with and without porous media. We use the empty cell as a reference for mixing in the presence of diffusion.   The experiments without porous media were carried out in thin cells of thickness $d=2$mm. The experiments with porous media were carried out in thicker cells (12 mm) to have a sufficient number of grain diameters in the cell thickness to represent a Darcy scale set up. 
As porous material, we used Fluorinated Ethylene Propylene (FEP 100 X, Chemours\textsuperscript{\textcopyright}) grains of mean grain diameter 2.5 mm. This material has the advantage of being transparent in water and have a refractive index ($n=1.34$) very close to that of water ($n=1.33$), limiting light scattering (see Appendix B for details) \cite{Amini2012}. 
These optical properties allow quantifying the depth integrated mixing width of the interface. The porosity of the packed FEP was measured to be 0.37. This is determined by calculating the volume of packed FEP beads based on the packed weight and the density of the FEP.

\subsection{Experimental protocol}
First, the cells are filled with deionized water. To fully saturate the media, we first injected CO$_2$ gas, which dissolved into the injected water hence preventing the presence of bubbles. The cells are placed on top of a blue LED panel to measure solute concentration by fluorescence. The mixing interface is produced by injecting clear and fluorescein tracer solutions at the same rate by a syringe pump. When the mixing front has stabilized, we image the interface with a digital camera (Sony A7s, F2.8/90) equipped with a green band pass filter (Midopt BN532). The image resolution was 0.04 mm per pixel in all experiments. Before each experimental run, we take two images of the cell fully filled with (i) pure water (image intensity $I_0$) and (ii) the fluorescein solution (image intensity $I_{max}$). A normalized tracer concentration is then computed with $c = (I-I_0)/(I_{max}-I_0)$, with $I$ the raw image intensity. Thus, $c$ varies between 0 and 1. 

Each experiment was performed for 9 flow rates, corresponding to 9 P\'eclet numbers (equation \eqref{eq:peclet}). The molecular diffusion coefficient of the fluorescein sodium salt is $4.2 \times 10^{-10}$ \si{m^2.s^{-1}}~\cite{Casalini2011}. In the non-uniform flow cell, the characteristic length scale of the velocity gradient $\gamma$ is estimated from the distance between the injection and the stagnation point $L_c=10$ cm for the setup without porous media. The compression rate is estimated by $v_{inj}/L$, with $v_{inj}$ the velocity at the injection and $L$ the distance between the injection and the stagnation point. The same characteristic length ($L_c=10$ cm) was used for the calculation of P\'eclet numbers. 
The 9 flow rates resulted in P\'eclet numbers ranging from 360 to 7150 and 930 to 18620 for the uniform and non-uniform flow cells respectively. 
The mixing width was estimated by fitting an error function \cref{eq:err} across the mixing interface at each distance ($x/d$). Because of residual light scattering by the FEP grains, the tracer concentration of the mixing front does not reach 0 in the side of the pure solution. Thus, the error function is fitted on a single side of the front, for which tracer concentrations are above 0.5 (\cref{figDispLog}a). For experiments with porous media, we triplicated each experiment by repacking the FEP grains to obtain results independent of specific grain arrangements. This lead to reproducible results (\cref{figDispLog}b), allowing us to capture the effect of flow non-uniformity on the evolution of the mixing scale.

\subsection{Experimental results}
\begin{figure}

\includegraphics[height=5.5cm]{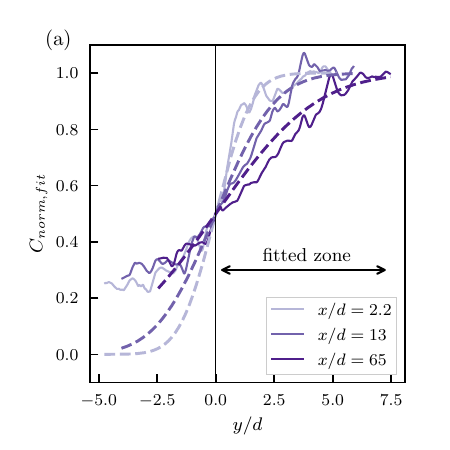}
\includegraphics[height=5.5cm]{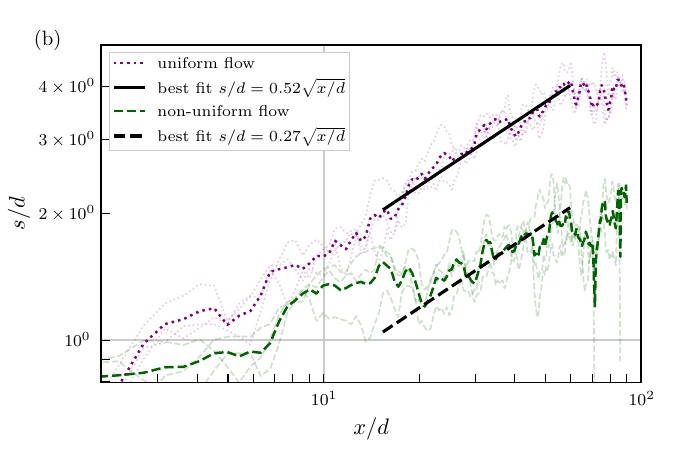}
        \caption{(a) procedure to obtain the mixing width by the fit of an error function on the upper part of the mixing interface (uniform flow, Pe~$=357$). The arrow indicates the fitted zone. (b) Average growth of the mixing width with distance in uniform (purple line) at Pe~$=5800$ and non-uniform flow (green line) at Pe~$=2320$.  The average is taken over 3 experiments with different grain packing. The light colored lines are the 3 replicates for each flow and the thick colored lines are their average. Best fits were estimated using a square root function ($s/d= A \sqrt{x/d}$) using the average mixing width of both flow cases for  $15<x/d<60$. }
        
        \label{figDispLog}
\end{figure}
\begin{figure}
    \begin{center}
        \includegraphics[width=\linewidth]{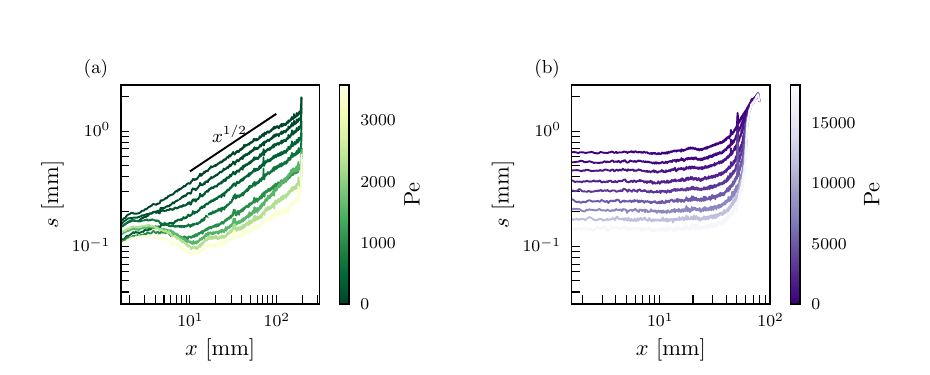}   
    \end{center}

    \caption{Mixing in Hele-Shaw cells without porous media in uniform versus non-uniform flows. (a) mixing width as a function of $x/U_x$ in the uniform flow configuration. (b) mixing width as a function of $x$ in the non-uniform flow configuration. Lines get darker when the P\'eclet number decrease.}
    \label{figDiff}
\end{figure}

Figure~\ref{figexpImages}b shows the experimental images obtained in the uniform and non-uniform flow cells with and without porous media. 
In absence of porous media and in uniform flows (\cref{figDiff}a), the transverse mixing interface follows the classical diffusive scaling $\sqrt{x}$ (table \ref{tab:width}). In contrast, when imposing non-uniform flows with constant compression (\cref{figDiff}b), the mixing width becomes constant along the interface. This highlights the balance between compression and diffusion, leading to the Batchelor scale (equation \eqref{eq:batchelor}, table \ref{tab:width}).  Note that the sudden increase observed for $x>5$ cm is caused by the finite size of the non-uniform flow cell, which induces a decrease of the compression rate at the outlet boundaries, and an increase of the mixing width.

In the presence of porous media, the mixing front dynamics are different as the mixing scale increases as $\sqrt{x/d}$ for both uniform and non-uniform flows (\cref{figDispLog}b). This is consistent with theoretical predictions (\ref{tab:width}). 
The non-uniform flow slightly depart from this tendency for $x<20d$ and $x>60d$ range. Indeed, close to the stagnation point, $x<20d$, the asymptotic theory (\cref{eq:dispSaddleWidth}) is not valid anymore. Close to the outlet, the mixing width starts to interact with the boundary of the cell and fluid compression is not  constant anymore, thus slowing down the observed growth. This limitation is also observed in the absence of porous media (see  \cref{figDiff}b). We thus limit the quantification to the range $20<x/d<60$. 
The estimated prefactor is 0.52 for uniform flow, corresponding to a transverse dispersivity of $\alpha_T=\sqrt{0.52/2}d=0.51 d$. For non-uniform flow, the pre-factor is 0.27, which is $\sqrt{3.7}$ times smaller than for the uniform flow, close to the predicted ratio of $\sqrt{3}$ (equation \eqref{eq:dispSaddleWidth}). 
This confirms that fluid compression sharpens mixing fronts in porous media, hence enhancing mixing rates.

\begin{figure}
    \begin{center}
        \includegraphics[width=0.7\linewidth]{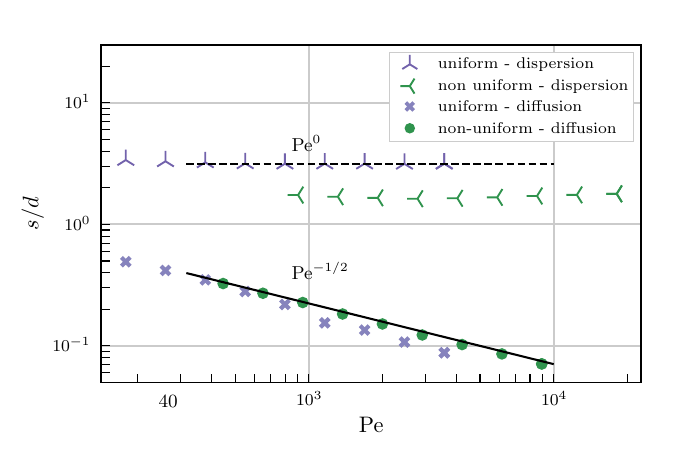}   
    \end{center}

    \caption{Mixing width as a function of P\'eclet for diffusion and dispersion. The mixing widths in uniform and non-uniform flows in porous media (with dispersion) and in uniform diffusive flow are plotted at 40$d$ (8 cm). For non-uniform diffusion, mixing width is averaged between the stagnation point and  10$d$ (2 cm) (see \cref{figDiff}). }
    \label{figDisp}
\end{figure}

In \cref{figDisp}, we plot the dependence of the mixing width with P\'eclet number at a given position along the front. In the presence of porous media, the mixing width is independent of P\'eclet number for both uniform and non-uniform flows. 
In contrast, in the absence of porous media, the mixing width decreases as Pe$^{-1/2}$ in uniform and non-uniform flow. These results are in agreement with the  theoretical predictions (table \ref{tab:width}). They  highlight the properties of the Batchelor scale, as a result of the balance between compression and diffusion, and its break down under dispersion at Darcy scale. The mixing scale is still systematically smaller under fluid compression and dispersion, leading to enhanced mixing rates (table \ref{tab:width}).  

\section{Discussion and conclusion}
We investigated the effect of local dispersion versus diffusion on the properties of mixing fronts developing in uniform and non-uniform flow. For the latter, we focused on flow topologies formed around a stagnation point in converging flows, a configuration that is common for mixing fronts in hydrological and hydrogeological systems.  We derived analytical solutions for the mixing width, concentration gradient and mixing rate for uniform and non-uniform flows and both diffusion and dispersion. We performed millifluidic tracer experiments in Hele-Shaw cells with and without porous media, showing good agreement with the theory. 
The results in the absence of porous media are representative of steady mixing fronts at pore scale, in a fracture or in open low Reynolds number flows. They show that fluid compression leads to the development of a fixed mixing scale independent of position but function of the P\'eclet number, hence highlighting the properties of the Batchelor scale. 
The results obtained with porous media are representative of Darcy scale mixing fronts.
In this case, the mixing scale grows with distance along the mixing interface but is independent of P\'eclet number, both for uniform and non-uniform flows. Fluid compression sharpens mixing fronts by enhancing concentration gradients by a factor $\sqrt{3}$, independent of the P\'eclet number. This leads to enhanced mixing, as known from theories considering diffusion and compression, but the scaling with distance and P\'eclet number differ from these classical predictions. 

These findings thus highlight the importance of considering  dispersion or diffusion  when studying mixing fronts in heterogeneous flows. 
The constant compression rate obtained locally in the considered non-uniform flow is analogous to chaotic flows, where a constant compression rate is sustained globally by repeated stretching and folding. This phenomenon is known to occur naturally at pores scale \cite{Lester2013, Lester2016, Heyman2020, Souzy2020} and can be created by engineered pumping and injection \cite{mays2012plume,trefry2012toward,neupauer2014chaotic,bertran2023enhancing}. The obtained scaling laws (table \ref{tab:width}) and impact of diffusion/dispersion are thus expected to be relevant for such chaotic flows.
An interesting perspective of this study is to investigate the impact of local dispersion on reaction rates in mixing fronts. 

\clearpage

\section*{Acknowledgment}
J.H. and G.R. acknowledge the grant ANR-19-CE01-0013.  J.H. acknowledges the ERC grant 101042466.
We thank E. Villermaux for fruitful discussions in the fifth summer school on Flow and Transport in Porous and Fractured Media at IESC, Carg\`ese.

\section*{Open Research}
All experimental images and the OpenFoam solver are available on the online repository Zenodo \cite{Izumoto2023b}.

\appendix

\section{Validation with simulations}
\label{sec:valid}
\begin{figure}
    \begin{tabular}{cc}
        \includegraphics[height=4.5cm]{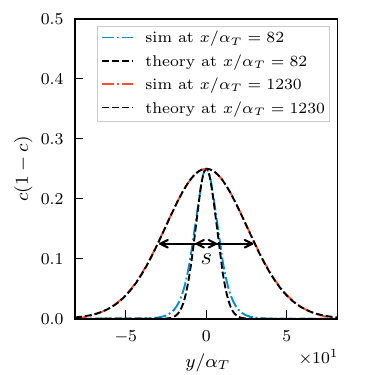}  &\includegraphics[height=4.5cm]{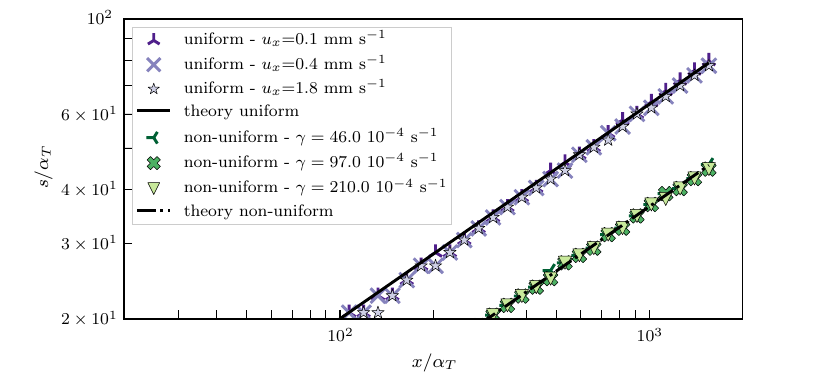}
    \end{tabular}
  \caption{Left: Simulated against predicted $c(1-c)$ profile for the non-uniform flow case taken at two distances from the injection. Right: normalized mixing interface width from simulation at different flow conditions against the theoretical $s/\alpha_T$ expression as given in \cref{eq:coWidth} and \cref{eq:dispSaddleWidth}. }
    \label{fig:simvsth}
\end{figure}

The 2D advection-dispersion \cref{eq:advdisp} is numerically solved using the software OpenFoam. \Cref{fig:sketch} provides examples of the computed concentration fields for uniform and non-uniform flows for diffusive and dispersive, where we highlighted the mixing zone by visualizing $c$. For the validation with the simulations, we compute the product $c (1-c)$ across the mixing interface, which follows \cref{eq:err}:
\begin{equation}
    c(1-c) =1/4 \left[ 1  - \erf \left( \frac{y}{ s} \right) \right]^2
    \label{eq:c(1-c)}
\end{equation}
with $s=2 \sqrt{\alpha_T x}$ for uniform and $s=2\sqrt{\alpha_T x/3}$ for non-uniform flow. Following \cite{de2022subsurface}, the width of the mixing zone is taken as the distance at half the maximum of $c(1-c)$. The growth of mixing width with distance is in \cref{fig:simvsth}, together with theoretical predictions.

\section {Transparency of Fluorinated Ethylene Propylene (FEP) grains}
\begin{figure}
    \centering
    \includegraphics[width=\linewidth]{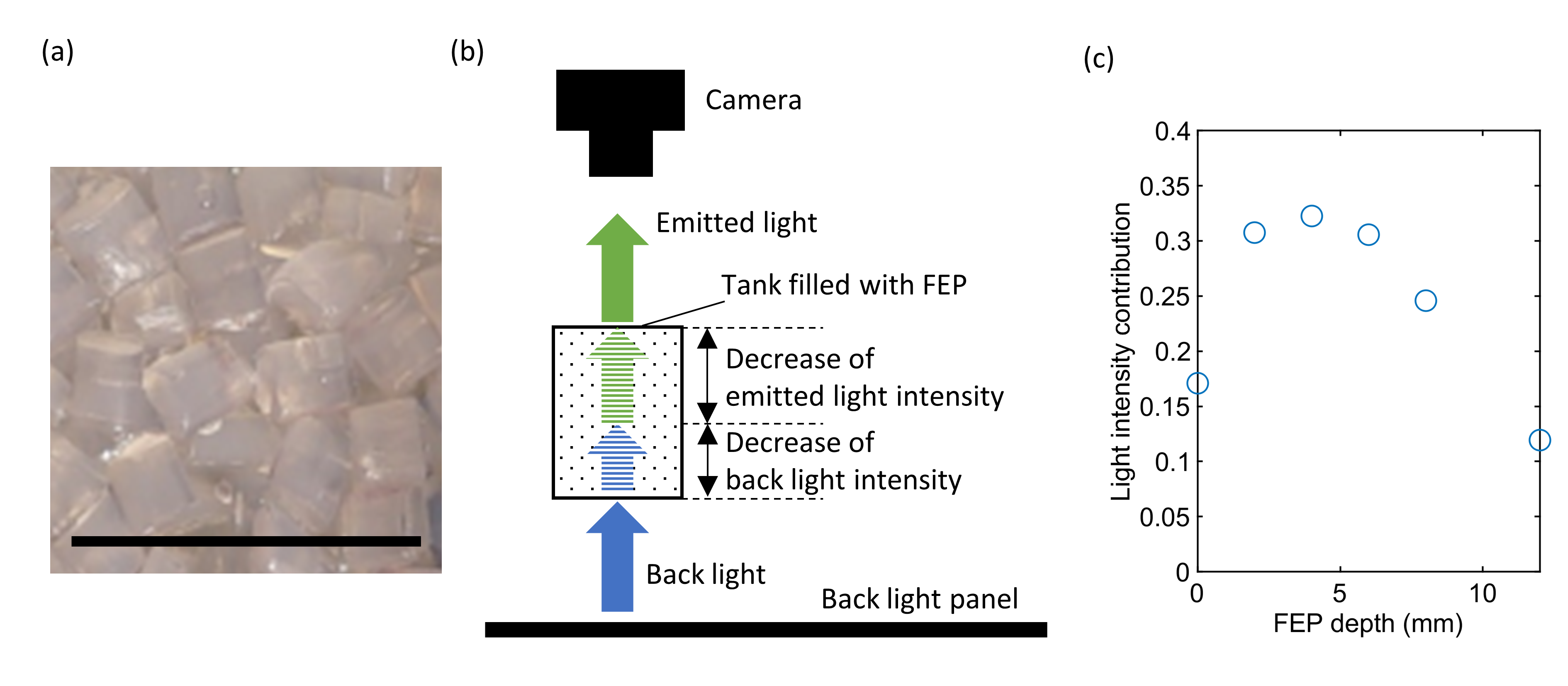}
    \caption{(a) Fluorinated Ethylene Propylene grains. The black line indicates 10 mm. (b) Schematic illustration of the emission of the light from fluorescein solution at a certain depth in a semi-transparent media. (c) Normalized received light intensity as a function of FEP packing thickness.}
    \label{figTransparency}
\end{figure}

The Fluorinated Ethylene Propylene (FEP 100 X, Chemours\textsuperscript{\textcopyright}) is a translucid plastic which has the advantage to have a index of refraction (1.34) close to water (1.33).
Since the FEP grains are not totally transparent for blue and green wavelengths (\cref{figTransparency}a), we investigated how much each depth contributes to the image intensity. We investigate (1) how far the blue excitation wavelengths penetrate into the media and (2) how much green light reemitted by the fluorescein reached the camera (\cref{figTransparency}b). To obtain these, we measured 1) the transparency of FEP to the blue backlight, 2) the transparency of FEP to the green light emitted from fluorescein, and 3) the transparency of fluorescein sodium salt solution to blue backlight. For 1) and 2), we placed different depth of the packed FEP (0, 2, 4, 6, 8, 12 mm) above the backlight (or above the illuminated fluorescein). The image intensities (normalized by the image intensity of 0 mm FEP depth) can be obtained as a function of FEP depth as $I_{F,b}(x_i)$ and $I_{F,f}(x_i)$ for packed FEP above the backlight and illuminated fluorescein, respectively, where $x_i$ is the depth of the FEP ($x_1=0 $ mm and $x_6 = 12$ mm). For 3), we placed different depth of fluorescein sodium salt solutions (0, 2, 4, 6, 8, 12 mm) above the backlight, which gave the normalized image intensity $I_{f,b}(x_i)$, where $x_i$ is the depth of the fluorescein sodium salt solution ($x_1=0$ mm and $x_6 = 12$ mm). The backlight reaches at certain depth ($I_b(x_j)$) can be calculated by considering the transparency of FEP and fluorescein solution to the backlight, and the porosity of the packed FEP $\phi = 0.373$ as:
\begin{equation}
\label{eqFEP}
    I_b(x_j) = I_b(x_{j-1}) + (I_{F,b}(x_j)-I_{F,b}(x_{j-1})) + \phi \times (I_{f,b}(x_j)-I_{f,b}(x_{j-1}))
\end{equation}
where $x_1$ corresponds to the bottom (0 mm depth) and we set $I_b(x_1)$ as 1. The decrease of the light intensity emitted by the fluorescein as a function of FEP depth is $I_{F,f}(x_i)$. Therefore, the contribution to the image intensity from a certain depth $I_c(x_j)$ is given by multiplying the backlight reached at certain depth and the decrease of the emitted light intensity as:
\begin{equation}
\label{eqFEP2}
    I_c(x_j) = I_b(x_j) \times I_{F,f}(x_{7-i})
\end{equation}
The results show that the $I_c$ takes maximum at certain depth (\cref{figTransparency}c). Nevertheless, maximum $I_c$ and minimum $I_c$ differs only by a factor two, which suggests that the image captured the mixing in entire depth even though the FEP grains are not perfectly transparent.

\bibliography{mixingdispersivemedia_WRR_for_arXiv.bib}

\end{document}